# Correlation between magnetooptical and transport properties of Sr doped manganite films

Yu.E. Samoshkina [a*], I.S. Edelman [a], M.V. Rautskii [a], M.S. Molokeev [a, b]

[a] *Kirensky Institute of Physics, Federal Research Center KSC SB RAS, 660036, Krasnoyarsk, Russian Federation*
[b] *Siberian Federal University, 660041, Krasnoyarsk, Russian Federation*
[*] uliag@iph.krasn.ru

The features of electronic structure of $La_{0.7}Sr_{0.3}MnO_3$, $Pr_{0.8}Sr_{0.2}MnO_3$, and $Pr_{0.6}Sr_{0.4}MnO_3$ polycrystalline films of different thickness have been investigated using magnetic circular dichroism (MCD) in the range of 1.1-4.2 eV. The temperature behavior of the samples electrical resistance were also has been studied. It was found that films with high Sr content (0.3 and 0.4) act as high-temperature semiconductors, while the maximum in the temperature dependences of these films resistivity indicates the transition of the samples to the metallic state at some temperature $T_{M-S}$, which is different for different sample thickness. The films with the lower Sr content (0.2) act as insulators in the used temperature range. The MCD spectra have been decomposed to the Gaussian-shaped lines, and the temperature dependence of intensity of each line has been analyzed in comparison with temperature dependence of the films magnetization and with their electric conductivity type. Different temperature behavior of the intensity of four specified Gaussian-lines was revealed for semiconductor films. In the case of insulating $Pr_{0.8}Sr_{0.2}MnO_3$ samples, the intensity of three specified Gaussian lines changes with the temperature in the same way as the magnetization changed. It was established that the lanthanide (La, Pr) type does not affect the MCD spectra shape for the films with the same electrical conductivity type. Besides, the correlation between the MCD data of the films and their conductivity type was revealed. Due to the detailed analysis of the specified Gaussian lines with taking into account the well-known in the literature absorption bands, lying outside the studied spectral region, the MCD bands for the studied manganite films have been identified with electronic transitions of a different nature.

**Keywords:** Thin films, Manganites, Crystal structure, Electrical transport, Electronic properties, Magnetic circular dichroism

# 1. Introduction

The substituted $RE_{1-x}A_xMnO_3$ manganites, where RE is a trivalent lanthanide and A is a divalent alkaline-earth metal (Ca, Sr, Ba et al.), are complex oxides whose magnetic behavior according to C. Zener [1, 2] is due to double exchange interaction between $Mn^{3+}$ and $Mn^{4+}$ ions. In addition, the manganites are systems with strong correlations between charge, spin, orbital, and lattice degrees of freedom, which leads to their rich phase diagrams [3, 4]. At present, the substituted manganites are payed great attention as research objects in the modern solid state physics. Particular interest to these materials is associated with unique combination of their properties, such as colossal magnetoresistance effect [5-7], magnetocaloric effect [8, 9], charge ordering [10, 11] and spin-polarized tunneling of charge carriers [7, 12]. However, despite the active study of the manganites, the picture of magnetic and electronic states determining macroscopic properties of the samples is not clear, and in order to describe it more fully, theoretical methods are still being searched [13].

Magneto-optical properties of substituted manganites in the range of 1-5 eV are extremely interesting as an additional source of information about electronic structure features of compounds. Numerous works deal with the magneto-optical Kerr Effect (MOKE) observed in the reflected light for the $La_{1-x}Sr_xMnO_3$ manganites (e.g., [13-17]). However, the interpretation of the MOKE spectra in terms of electronic transitions is not a trivial task, since the total rotation of these effects is a complex function of the diagonal and off-diagonal components of the permittivity tensor. The results of MOKE measurement depend critically on the quality of the samples surface. Discrepancies observed between MOKE spectral characteristics for $La_{1-x}Sr_xMnO_3$ (LSMO), namely, the change in the spectrum shape with the change of the sample thickness [14] and concentration of the $A^{2+}$ alkaline-earth metal [15-17] can be associated with the peculiarities of the MOKE measurement. In addition, the dependence of MOKE spectrum shape on the type of rare earth was not discussed in the literature. The temperature dependences of different magneto-optical bands were not investigated before. Thus, there is no information about comparison of the temperature behavior of the magnetization and the magneto-optical bands in the same sample.

Compared to MOKE, the magnetic circular dichroism (MCD) effect is more informative since it is observed directly at the electron transition frequency and described by the dissipative line shape. The MCD effect is the difference between the absorption coefficients of the right and left polarized (relatively to the magnetic vector direction) light waves (Δk). In order to obtain the expression describing the MCD effect in a homogeneous ferromagnetic, it is necessary to present the complex Faraday rotation (FR) (1) as a sum of real and imaginary parts

$$\tilde{\alpha}_{FR} = b(\Delta n - i\Delta k), \qquad (1)$$

where $b$ is a coefficient, $\Delta n$ is a difference between the refractive coefficients, and $\Delta k$ is a difference between the absorption coefficients. The chain of simple transformations (2-3) with taking into account $\varepsilon'_{xx} = n^2 - k^2$ and $\varepsilon''_{xx} = 2nk$ gives an expression for the MCD (4). $\varepsilon_{xx}$ and $\varepsilon_{xy}$ are a diagonal and off-diagonal permittivity tensor components, respectively.

$$\Delta n - i\Delta k = \frac{\varepsilon_{xy}}{\sqrt{\varepsilon_{xx}}} \quad (2)$$

$$(\Delta n - i\Delta k)^2 = \left(\frac{\varepsilon'_{xy} - i\varepsilon''_{xy}}{\sqrt{\varepsilon'_{xx} - i\varepsilon''_{xx}}}\right)^2 \quad (3)$$

$$\Delta k = \frac{n\varepsilon''_{xy} - k\varepsilon'_{xy}}{n^2 + k^2} \quad (4)$$

However, it is worthy to note that MCD was seldom used for the study of manganites. In particular, in Ref. [18] MCD in the epitaxial LSMO (x=0.3) films was studied depending on the temperature and the substrate type. Nevertheless, in Ref. [18] MCD was not considered in terms of electronic transitions. Recently, we investigated the MCD spectra for the polycrystalline LSMO (x=0.3) films [19] and $Pr_{1-x}Sr_xMnO_3$ (PSMO, x=0.4 and 0.2) films [20]. In this work, we perform a comparative study of the MCD data for the LSMO (x=0.3) and PSMO (x=0.4 and 0.2) samples and revise the nature of the spectra features. The primary focus is on effect of the lanthanide type and the electrical conductivity type of the samples on the MCD spectra characteristics.

## 2. Material and methods

The films were prepared by the dc magnetron sputtering with the "facing-target" scheme [21]. This scheme allows transferring evaporating material from the target to the substrate without changes in the composition. The stoichiometric single phased LSMO (x=0.3), PSMO (x=0.4) and PSMO (x=0.2) targets were fabricated by the solid-state synthesis from the stoichiometric $La_2O_3$ or $Pr_2O_3$, SrO, and $MnO_2$ powders. Before sputtering, the residual pressure in the vacuum chamber was $3 \times 10^{-6}$ Torr. The operating pressure of the Ar and $O_2$ gas mixture (4:1) was $3 \times 10^{-3}$ Torr. Thin films were grown on the Y-oxide stabilized zirconium dioxide (YSZ) substrates. The substrate temperature during sputtering was 750 °C. The film thickness, d, varied for LSMO (x=0.3) from 20 to 90 nm, for PSMO (x=0.4) from 50 to 130 nm, and for PSMO (x=0.2) from 50 to 150 nm, it was controlled by the deposition time and determined ex-situ by the X-ray fluorescence analysis. The analysis of the samples using the Rutherford backscattering method showed that the chemical composition of the obtained films corresponded to the stated above stoichiometry.

The investigation of magnetic properties of the studied films in Refs. [19, 22] showed that the Curie temperature ($T_C$) was ~ 300 K for LSMO (x=0.3), ~ 250 K for PSMO (x=0.4), and ~ 120 K for PSMO (x=0.2). Tc value is lower for films in comparison with their bulk analogues, which can be explained by the displacement of the phase separation boundaries caused by the strain between the lattices of individual crystallites [23].

The crystal structure and phase purity of the samples were examined by the X-ray diffraction (XRD). The XRD data were collected on the Bruker D8 ADVANCE diffractometer using CuKa radiation at 40 kV and 40 mA in a step mode (the step size $0.016^0$ and 10 s counting time per step) over the range from 10 to $140^0$. The X-ray beam was controlled by a 0.6 mm fixed slit. Reflections from the substrate were simulated by individual peaks with the Pearson VII profile. The other peaks were accounted by the corresponding phase using Rietveld refinement. The Rietveld refinement was performed using the program TOPAS 4.2. PSMO (x=0.3) was taken as the original structure [24]. Phases LSMO (x=0.3), PSMO (x=0.4), and PSMO (x=0.2) were obtained from it by Pr→La replacement and changing occupancy of Pr/Sr.

Electrical resistivity of the films was measured in a DC (direct current) mode at fixed value of the current using the standard four-point probe technique with an original facility based on a helium cryostat, an electromagnet and a precise Keithley 2400 Source Meter. Current was directed in the films plane. Temperature dependences were measured in the range of 70-300 K and magnetic field up to 3.5 kOe. Magnetic field was applied in two orientations: along and normally to the sample plane.

The MCD spectra were measured using the modulation of the light wave from the right-hand to the left-hand circular polarization relatively to the applied magnetic field direction. The modulator was made of a fused silica prism with a glued piezoelectric ceramic element. The MCD measurement technique is described in more detail in Ref. [25]. It should be noted that magnetic field (H) must have the value sufficient to magnetize a sample to saturation in the chosen direction. Because of the limited H value, this condition was not fulfilled while measuring MCD in the LSMO (x=0.3) films at low temperatures, and thus, the MCD temperature dependence was not correct. To overcome this difficulty, we used the oblique incidence of the light beam on the sample ($12^0$) that provided the magnetic saturation of the sample in its plane throughout the temperature range used. In this case, the MCD value was determined by the sample magnetization component along the light beam. In the case of the PSMO (x=0.4 and 0.2) films, the MCD was measured in the normal geometry: the magnetic vector and the light beam were directed normally to the sample plane. The MCD in both cases was measured in the spectral range of 1.1-4.2 eV in the magnetic field up to 3.5 kOe in the temperature range of 100-300 K. The measurement accuracy was about $10^{-4}$, and the spectral resolution was 0.002-0.006 eV depending on the wavelength. Temperature was measured with accuracy 1 K.

## 3. Results and Discussion

### 3.1. Films structure

The XRD patterns for the thickest LSMO (x=0.3), PSMO (x=0.4), and PSMO (x=0.2) films showed large peaks corresponding to the YSZ substrate and a number of relatively narrow peaks with small intensity corresponding in each case to only one phase. The crystal structure of the studied samples was refined within the orthorhombic Pnma space group with lattice parameters and average crystallites size presented in Tables 1-3. Structural parameters of the polycrystalline films are in agreement with the structural data for the LSMO (x=0.3) and PSMO (x=0.4 and 0.2) samples reported earlier in Refs. [26-34]. It should be noted that the pronounced texture was revealed in the LSMO (x=0.3) films, while for the PSMO (x=0.4 and 0.2) films the texture was not observed. In addition, we should clarify that LSMO (x=0.3) has rhombohedral R-3c structure in bulk form [35], but in nanoscale samples both rhombohedral [36] and orthorhombic [26-28] structures can form.

Table 1. Structural parameters and average crystallites size of the LSMO (x=0.3) film, d = 90 nm.

| LSMO (x=0.3) | This work | Ref. [26] | Ref. [27] | Ref. [28] |
|---|---|---|---|---|
| | Thin films | Thin films | Wires | Polycrystalline nanomaterials |
| Space group | Pnma | Pnma | Pnma | Pnma |
| Substrate | YSZ (001) | SrTiO3 | - | - |
| a (Å) | 5.51 (5) | 5.480 | 5.454 | 5.446 |
| b (Å) | 7.70 (1) | 7.809 | 7.695 | 7.707 |
| c (Å) | 5.50 (5) | 5.483 | 5.458 | 5.483 |
| V (Å$^3$) | 233.35 (4) | 234.64- | 229.0.64 | 230.175 |
| Average crystallites size (nm) | 95 (8) along a and/or c axes | - | - | - |
| Coefficient texture | 0.18 (6) | - | - | - |

Table 2. Structural parameters and average crystallites size of the PSMO (x=0.4) film, d = 130 nm.

| PSMO (x=0.4) | This work | Ref. [29] | Ref. [30] | Ref. [31] |
|---|---|---|---|---|
| | Thin films | Single crystal | Polycrystalline sample | Powder sample |
| Space group | Pnma | Pnma | Pnma | Pnma |
| Substrate | YSZ (311) | - | - | - |
| a (Å) | 5.4465 (7) | 5.4859 (1) | 5.447 (4) | 5.4460 (2) |
| b (Å) | 7.721 (1) | 7.6790 (1) | 7.684 (5) | 7.6796 (3) |
| c (Å) | 5.422 (1) | 5.4443 (9) | 5.488 (4) | 5.4892 (2) |
| V (Å$^3$) | 228.00 (6) | 229.348 | 229.8 (5) | 229.575 (1) |
| Average crystallites size (nm) | 45 (7) along b axis 62 (3) along a and c axes | - | 113 | - |

Table 3. Structural parameters and average crystallites size of the PSMO (x=0.2) film, d = 150 nm.

| PSMO (x=0.2) | This work | Ref. [32] | Ref. [33] | Ref. [34] |
|---|---|---|---|---|
| | Thin films | Powder sample | Particles | Polycrystalline sample |
| Space group | Pnma | Pnma | Pnma | Pnma |
| Substrate | YSZ (311) | - | - | - |
| a (Å) | 5.480 (1) | 5.4846 | 5.485 | 5.47 |
| b (Å) | 7.761 (1) | 7.7376 | 7.784 | 7.72 |
| c (Å) | 5.462 (5) | 5.5056 | 5.479 | 5.48 |
| V (Å$^3$) | 232.3 (2) | 233.645 | 233.92 | 231.412 |
| Average crystallites size (nm) | 31 (3) along b axis  54 (4) along a and c axes | - | - | - |

### 3.2. Magnetic circular dichroism

Typical MCD spectra obtained for the LSMO (x=0.3) and PSMO (x=0.4 and 0.2) films at different temperatures are shown in Figs. 1 and 2. Two main features of the same sign characterize the MCD spectra shape of the studied samples: an intense broad band in the region of 3.2 - 3.4 eV and a weaker asymmetric broad band in the region of 1.7-2 eV. The intense broad band at 3.3 eV was observed also in the MCD spectra for the LSMO (x=0.3) films in Ref. [18]. As the temperature decreases, the intensity of both MCD bands increases until the lowest temperature used. In addition, in the case of the LSMO (x=0.3) and PSMO (x=0.4) films, specific MCD value behavior is observed in the region of 2.2-2.5 eV: at some temperature $T_s$, the MCD value changes its direction (see inserts in Figs. 1 and 2a). Then, with further temperature decrease, the relatively weak band of the positive sign appears near 2.3-2.4 eV in the MCD spectra for the LSMO (x=0.3) and PSMO (x=0.4) films. In the case of the PSMO (x=0.2) films, a usual temperature variation of the MCD value in this region is observed and an additional band of the positive sign does not arise (Insert in Fig. 2b). The film thickness does not affect the MCD spectra shape. However, $T_s$ for the LSMO (x=0.3) and PSMO (x=0.4) films decreases with the samples thickness reduction.

### 3.3. Decomposition of the experimental MCD spectra

Since the observed asymmetry of the magneto-optical bands can be caused by a superposition of several absorption lines, the MCD spectra of the samples were decomposed into several Gaussian-shaped lines for each temperature. The decomposition was performed on the minimal number of lines. The Gaussian line amplitude (Δk), position (E), and line width at the half-height (ΔE) were the fitting parameters. In the case of the PSMO (x=0.2) films, we obtained a good fit of the sum of three Gaussian lines with the experimental spectra for all temperatures studied. In the case of the LSMO (x=0.3) and PSMO (x=0.4) films, good enough fittings with the experimental spectra were obtained for the sum of four the Gaussian lines. A typical example is shown in Fig. 3. Positions of the Gaussian lines are presented in Table 4 (in cm$^{-1}$ and in eV for the convenience in case of comparing with data of other authors). The shift of the Gaussian lines positions to lower energies with increasing Sr content in the samples is observed in this Table. The shift is larger when coming from x=0.2 to x=0.3 in comparison to the transition from x=0.3 to x=0.4.

Table 4. Positions (E) of the Gaussian lines in the decomposed MCD spectra for the LSMO (x=0.3) and PSMO (x=0.4 and 0.2) samples. The error of E position with temperature change is +/- 0.03 eV.

| Sample | $E_1$ | | $E_2$ | | $E_3$ | | $E_4$ | |
|---|---|---|---|---|---|---|---|---|
| | cm$^{-1}$ | eV | cm$^{-1}$ | eV | cm$^{-1}$ | eV | cm$^{-1}$ | eV |
| PSMO (x=0.4) | 11260 | **1.40** | 14105 | **1.75** | 18880 | **2.34** | 25593 | **3.18** |
| LSMO (x=0.3) | 11721 | **1.46** | 14850 | **1.84** | 19170 | **2.38** | 25827 | **3.21** |
| PSMO (x=0.2) | 12998 | **1.61** | 16291 | **2.02** | - | - | 26889 | **3.34** |

### 3.4. Resistivity temperature dependence

Temperature dependences of the resistivity for the LSMO (x=0.3) and PSMO (x=0.4 and 0.2) films are shown in Fig. 4. The films with high Sr content (0.3 and 0.4) act like high-temperature semiconductors. With the temperature decrease, one can observe the maximum in the resistivity curves for these samples (Figs. 4 a and 4b), indicating the transition to the metallic state. It is worthy to note that the film thickness affects on the resistivity value and the temperature of the metal-semiconductor transition, $T_{M-S}$. As the film thickness decreases, the resistivity value increases and $T_{M-S}$ shifts toward low temperatures. Magnetic field application leads to the decrease in resistivity, while the temperature $T_{M-S}$ shifts insignificantly toward higher temperatures. The films with low Sr content (0.2) act as insulators in the whole temperature range (Fig. 4c). In addition, the samples resistivity behavior does not change when the magnetic field up to 3.5 kOe is applied. The electrical resistivity data of the films are in good agreement with the phase diagrams of their bulk analogs [3, 4].

The dependence of $T_{M-S}$ temperature on the film thickness can be explained, taking into account the samples structure. It is well known that physical properties of polycrystalline films are sensitive to the lattice strain and grain boundaries [37-39]. Furthermore, transition temperature $T_{M-S}$ is sensitive to grain boundaries only, since a significantly high scattering on the grain boundaries leads to enhanced resistivity and reduction in the $T_{M-S}$ [37]. Thus, the $T_{M-S}$ shift toward low temperatures, observed in Figs. 4a and 4b, is explained by the crystallites size decrease with reduction in the films thickness and growth of boundary effects.

### 4. Discussion

The MCD spectra shown in Figs. 1 and 2 for the studied LSMO and PSMO films are similar in shape to the MOKE spectra observed earlier for the epitaxial LSMO (x=0.3) films in Refs. [15-17]. However, in the cited works, different changes in the MOKE spectrum shape were observed when the film thickness changed. Thus, interpretations of spectral features in the off-diagonal permittivity tensor components calculated for LSMO (x=0.3) from the MOKE spectra varied in different works and still remain controversial. It is also worth noting that due attention was not paid to the spectral band of positive sign near 2.5-2.6 eV.

As noted above, the MCD effect is observed directly on the absorption bands. Besides, we established that the samples thickness does not affect the MCD spectra shape of the LSMO (x=0.3) and PSMO (x=0.4 and 0.2) films. Thus, the MCD spectra decomposition into Gaussian-shaped lines and analysis of the temperature dependence of the specified lines intensity provide an opportunity for investigating the electronic states in the samples directly. (Next in the text $E_1$-$E_4$ Gaussian lines are identified as $E_1$-$E_4$ MCD bands.) The intensity temperature dependences of the MCD bands obtained by the decomposition are presented in Figs. 5a, 5b, and 5c for the thickest LSMO (x=0.3) and PSMO (x=0.4 and 0.2) films, respectively. Since MCD temperature dependence reflects the temperature dependence of the sample magnetization, the temperature behavior of the films magnetization published in Refs. [19, 22] are also presented in Fig. 5.

As it can be seen in Figs. 5a and b, different temperature dependences of the MCD bands intensity are intrinsic to the semiconductor films. One can highlight the following main points. 1) As the temperature decreases, the $E_1$, $E_2$ and $E_4$ bands intensity increases and their temperature behavior is similar with each other up to some temperature after which the $E_1$ and $E_2$ bands experience a kink (Figs. 5a and b). Then the intensity of $E_1$ and $E_2$ bands decreases with the temperature reduction. 2) The intensity of $E_3$ band increases when the temperature decreases (Figs. 5a and b), however, it obeys another law.

Comparison of the temperature behavior of the $E_1$-$E_4$ bands intensity with the temperature dependence of the semiconductor films magnetization (curves M in Figs. 5a and 5b) is not a simple task. It is known that the temperature behavior of the film magnetization is sensitive to the sample structure [39]. Thus, an observed discrepancy between M temperature dependences of two semiconductor films can be explained by the effect of structural features of the samples. In the case of LSMO (x=0.3), crystallite sizes are larger and they are highly textured in the film plane, while for PSMO (x=0.4), crystallites are significantly smaller and they are oriented randomly in the films plane (Table 1 and 2, respectively).

At the same time, it should be noted that, in the case of the insulating PSMO (x=0.2) films, the form of the temperature dependency of the intensity is identical for all specified MCD bands and corresponds to the temperature dependence of the samples magnetization (Fig. 5c).

In order to determine the nature of the considered magneto-optical bands, one should pay attention, primarily, to the optical spectra of undoped manganites. For the $RMnO_3$ samples (R=La, Pr, Nd, Gd, Tb) containing only $(Mn^{3+}O_6)^9$ complexes, two broad intense bands centered near 2.0 and 4-5 eV were observed in the spectra of the optical conductivity as well as the optical absorption [13, 40-42]. The nature of the band near 2 eV was discussed by various authors. In the Refs. [13, 40], which is devoted to the determination of the band structure of $LaMnO_3$, this band was attributed to the interband transitions between occupied and empty $e_g$ states belonging to $Mn^{3+}$. It is believed that such a transition is possible due to the hybridization of $e_g$ states for the manganese with $O_{2p}$ states. At the same time, the authors studying experimentally optical and magneto-optical spectra of the undoped manganites attributed the band near 2 eV to the spin-allowed crystal-field d-d transition in the $Mn^{3+}$ ions [43, 44]. The effect of manganites doping with Sr is the formation of $(Mn^{4+}O_6)^{8-}$ octahedral complexes along with $(Mn^{3+}O_6)^{9-}$ complexes that leads to the shift of the spectral weight of the optical conductivity to the low energy region [13, 41, 43, 44]. Such a shift was associated in Refs. [43, 44] with both the appearance of additional absorption bands formed by electron transitions in $Mn^{4+}$ ions and the screening of the crystal field by the hole density partly localized on surrounding oxygen ions. The band in the region of 3.2 - 3.4 eV was observed only in magneto-optical spectra of doped manganites [15-18, 43-45]. The nature of this band was usually associated by different authors either with the $Mn^{3+}e_g \rightarrow O_{2p} \rightarrow Mn^{4+}e_g$ charge transfer transitions [15, 16] or with the spin-allowed d-d transitions in $Mn^{4+}$ ions [43, 45].

In the MCD spectra of the studied LSMO and PSMO films three main bands ($E_1$, $E_2$, $E_4$) were determined at the energies presented in Table 4. According to the cluster model, which considers the effect of the crystal field only, one could associate these bands with the spin-allowed crystal-field d-d transition $^5E_g \rightarrow {}^5T_{2g}$ in the $Mn^{3+}$ ions ($E_1$ band) and d-d transitions $^4A_{2g} \rightarrow {}^4T_{2g}$ and $^4A_{2g} \rightarrow {}^4T_{1g}$ in the $Mn^{4+}$ ions ($E_2$ and $E_4$ bands, respectively). The energy positions of the specified bands are in good agreement with the Tanabe-Sugano diagrams for the $d^3$ ($Mn^{4+}$) and $d^4$ ($Mn^{3+}$) electronic configurations [46], taking into account different strength of the cubic crystal field in the $(Mn^{4+}O_6)^{8-}$ and $(Mn^{3+}O_6)^{9-}$ octahedral complexes.

The presence of the kink in the temperature dependence of the $E_1$ and $E_2$ bands intensity (Figs. 5a and 5b) for the LSMO (x=0.3) and PSMO (x=0.4) films can be explained by the spectral intensity redistribution between excitations of different nature in the region of 0-2 eV. Rearrangement of the optical conductivity and absorption spectra with the temperature decrease, observed in Refs. [41, 47, 48] for lanthanum and neodymium manganites, argues for this assumption. The spectral weight transfer from the higher energy (above 1 eV) to the lower energy (below 1 eV) occurred when these materials were cooled below their Tc and were in the metallic state. Different authors explain the observed spectra rearrangement in a different way. In Ref. [47] it was assumed that the dominant broad peak near 1.2 eV in the optical conductivity spectrum of the neodymium manganite is associated with the $Mn^{3+}e_g \to O_{2p} \to Mn^{4+}e_g$ charge transfer transition. At the same time, it was shown that the change in the spectrum at the temperature decrease agrees with the model, which includes both the double-exchange and the dynamic Jahn-Teller effect. In Refs. [41, 48] the optical band near 1.5 eV was associated with the interband transitions between the exchange-split conduction band consisting of the $e_g$ states (probably, with $Mn^{3+}$ ions participation only) hybridized strongly with the $O_{2p}$ states. Additionally, the spectral weight shift with temperature decrease was interpreted as a transition from the interband excitations to the intraband ones. In particular, in Ref. [41] there was noted that the intraband part contains incoherent ω-independent broad structure, and a sharp coherent Drude peak with anomalously low spectral weight. Besides, the authors pointed the complex nature of the low-energy edge of the optical spectra and suggested possible strong carrier scattering, dynamical Jahn-Teller effect, and an orbital degree of freedom in the $e_g$ conduction state. However, similar behavior was not observed in the optical conductivity spectrum of the insulating ferromagnetic LSMO (x=0.1) [41]. In the case of the insulating PSMO (x=0.2) films, there is also no kink in the temperature dependence of the $E_1$ and $E_2$ bands intensity. Thus, the obtained data indicate that there is correlation between optical (magneto-optical) properties of the manganites in the region of 1-2 eV and their conductivity type.

It is remarkable that the MCD band with the positive sign centered near 2.3-2.4 eV ($E_3$) also correlates with the conductivity type of the studied films. The $E_3$ band is not observed in the MCD spectra for insulating PSMO (x=0.2) but is typical for the LSMO (x=0.3) and PSMO (x=0.4) samples, which exhibit the transition to the metallic state at $T_{M-S}$ temperature (Figs. 2 and 4). Correlation data between the $E_3$ band and $T_{M-S}$ transition temperature are presented in Table 5 for the LSMO (x=0.3) and PSMO (x=0.4) films of different thickness. Despite the ambiguity of obtaining and interpreting the MOKE effect data, it is worth noting that a similar band of positive sign near 2.5-2.6 eV was observed earlier in a MOKE spectrum and a spectrum of the off-diagonal component of the permittivity tensor for the LSMO (x=0.3) [15-17]. In Ref. [15], in accordance with the band structure calculations, this band was associated with the $t_{2g} \to O_{2p} \to e_g$ interband transitions, probably, with the $Mn^{4+}$ ions participation. For the insulating LSMO with weak hole doping (x=0.07) there is no this band in the MOKE spectrum [43]. Thus, an appearance of the $E_3$ band in the magneto-optical spectra is associated obviously with a change in the band structure of the doped manganites, when the contribution of free charge carriers becomes predominant.

It is important to note that in the MOKE and MCD spectra for LSMO (x=0.3) three bands of positive sign are in the range of 1-5 eV [15-18], namely, one is in the range of 1-1.5 eV, second is in the range of 2.3-2.6 eV and third is in the range of 4-4.5 eV. Second of these bands is observed in the MCD spectra for the LSMO (x=0.3) and PSMO (x=0.4) films in the present study and defined as the $E_3$ band. If we apply literary data to the MCD spectra observed in the present study, then we can expect two positive MCD band, namely, one at the energy lower than $E_1$ and one at the energy higher than $E_4$. Taking into account different interpretations of the nature of optical and magneto-optical bands, which are presented in Refs. [13, 15, 40, 49], we can assume that the indicated positive sign bands are associated with the $Mn^{3+}e_{g1} \rightarrow O_{2p} \rightarrow Mn^{3+}e_{g2}$ interband transitions and the $Mn^{3+}e_g \rightarrow O_{2p} \rightarrow Mn^{4+}e_g$ charge transfer transitions, respectively. However, for the insulating LSMO (x=0.07 [43] and x=0.15 [50]) only one positive sign band in the range of 4-4.5 eV is characteristic, which is associated with the $Mn^{3+}e_g \rightarrow O_{2p} \rightarrow Mn^{4+}e_g$ charge transfer transitions. The $E_3$ band can be ascribed, most probably, to the $Mn^{4+}t_{2g} \rightarrow O_{2p} \rightarrow Mn^{4+}e_g$ interband transitions. Along with that, the discussion of the temperature dependence of the $E_3$ band intensity remains currently an open question. We can only draw attention to the fact that its course is similar to the temperature dependence of the spin polarization of a free LSMO (x ≈ 0.3) surface shown in Ref. [51].

Table 5. Correlation between MCD band near 2.3-2.4 eV and $T_{M-S}$ transition temperature for the LSMO (x=0.3) and PSMO (x=0.4) films. $T_s$ is the temperature at which the MCD value in the region of 2.2-2.5 eV changes its course.

| Samples | d (nm) | $T_s$ (K) | $T_{M-S}$ (K) |
|---|---|---|---|
| LSMO (x=0.3) | 20 | 150 | 155 |
|  | 50 | 190 | 188 |
|  | 90 | 210 | 197 |
| PSMO (x=0.4) | 50 | 145 | 154 |
|  | 80 | 150 | 158 |
|  | 130 | 200 | 171 |

Nature of all specified MCD bands and the possible additional bands of positive sign taken from Refs. [15-18] is presented in Table 6 for the studied LSMO and PSMO films. Finally, it is worth noting that the MCD spectra shape as well as temperature dependences of the MCD bands intensity are very close for the LSMO (x=0.3) and PSMO (x=0.4) films with different lanthanide ions and distinguish noticeably for the insulating PSMO (x=0.2) film. Thus, the lanthanide type does not affect the MCD spectra shape for LSMO (x=0.3) and PSMO (x=0.4) films with the same electrical conductivity type. The spectra for such samples vary at different Sr content, namely, the magneto-optical bands shift to the lower energies at higher Sr content in the sample. This behavior is quite natural and can be explained by the higher hole density in the sample.

Table 6. Nature of the specified MCD bands and of the possible additional (add.) bands of positive sign in the range of 1-5 eV for the LSMO (x=0.3) and PSMO (x=0.4 and 0.2) films.

| Sample | add. band (eV) | $E_1$ (eV) | $E_2$ (eV) | $E_3$ (eV) | $E_4$ (eV) | add. band (eV) |
|---|---|---|---|---|---|---|
| PSMO (x=0.4) | < 1.40 | 1.40 | 1.75 | 2.34 | 3.18 | > 3.18 |
| LSMO (x=0.3) | < 1.46 | 1.46 | 1.84 | 2.38 | 3.21 | > 3.21 |
| PSMO (x=0.2) | - | 1.61 | 2.02 | - | 3.34 | > 3.34 |
| Nature | $Mn^{3+}e_{g1} \to O_{2p} \to Mn^{3+}e_{g2}$ interband transitions | $^5E_g \to {}^5T_{2g}$ d–d transitions in $Mn^{3+}$ ions | $^4A_{2g} \to {}^4T_{2g}$ d–d transitions in $Mn^{4+}$ ions | $Mn^{4+}t_{2g} \to O_{2p} \to Mn^{4+}e_g$ interband transitions | $^4A_{2g} \to {}^4T_{1g}$ d–d transitions in $Mn^{4+}$ ions | $Mn^{3+}e_g \to O_{2p} \to Mn^{4+}e_g$ charge transfer transitions |

## 5. Conclusion

The crystal structure, spectral (in the range of 1.1-4.2 eV) and temperature (T >100 K) dependences of the MCD effect as well as the temperature behavior of the electrical resistance have been studied for the LSMO (x=0.3) and PSMO (x=0.4 and 0.2) polycrystalline films of different thickness. It was defined that structural parameters of the polycrystalline films agree with the structural data for the LSMO (x=0.3) and PSMO (x=0.4 and 0.2) samples reported earlier. It was found that the film thickness does not affect the MCD spectra shape, however it affects the resistivity value and the temperature of the metal-semiconductor transition, which is observed for the LSMO (x=0.3) and PSMO (x=0.4) samples.

The MCD spectra have been decomposed to the Gaussian-shaped lines, and the temperature dependence of intensity of each line has been analyzed in comparison with temperature behavior of the films magnetization and with their electric conductivity type. The correlation between MCD data of the films and their conductivity type was revealed. For the semiconductor LSMO (x=0.3) and PSMO (x=0.4) films different temperature dependencies of the intensity for four specified MCD bands (Gaussian lines) were established, and it was shown that the lanthanide type does not affect the MCD spectra shape. The spectra for these samples vary at different Sr content, namely, the MCD bands shift to the lower energies at higher Sr content in the film. This behavior was explained by the higher hole density in the sample. For the insulating PSMO (x= 0.2) films it was found that the intensity of three specified MCD bands changes with the temperature in the same way as the magnetization changes.

Three negative sign MCD bands, which is characteristic for all studied films, were determined and attributed to the spin-allowed d-d electron transitions in the $Mn^{3+}$ and $Mn^{4+}$ ions. The nature of the positive sign MCD band near 2.3-2.4 eV characteristic only for the LSMO (x=0.3) and PSMO (x=0.4) films was discussed and associated with the $Mn^{4+}t_{2g} \to O_{2p} \to Mn^{4+}e_g$ interband transitions. Nevertheless, the temperature dependence of this band intensity remains unclear. The kink presence in the temperature dependence of the intensity of the low-energy MCD bands detected also for the LSMO

(x=0.3) and PSMO (x=0.4) films was explained by the spectral intensity redistribution between excitations of different nature in the region of 0-2 eV. In addition, in the MCD spectra for the LSMO (x=0.3) and PSMO (x=0.4) films the existence in the low-energy and high-energy region of one positive sign MCD band associated with the $Mn^{3+}e_{g1} \rightarrow O_{2p} \rightarrow Mn^{3+}e_{g2}$ interband transitions and the $Mn^{3+}e_g \rightarrow O_{2p} \rightarrow Mn^{4+}e_g$ charge transfer transitions, respectively, was assumed.

Note that the observed correlation effects must also be traced on single crystal manganites, as justified by the MOKE data in the literature. The obtained results are an important step towards understanding of electronic states in substituted manganites and the search of new functional materials on their basis.

**Acknowledgments**

The authors are grateful to A.V. Malakhovskii (Kirensky Institute of Physics, Federal Research Center KSC SB RAS) for fruitful discussions of the magneto-optics data and V.I. Chichkov and N.V. Andreev (National University of Science and Technology MISiS) for the preparation of the manganite films. The work was supported partly by Russian Academy of Sciences, [grant number 0356-2017-0030], and the Russian Foundation for Basic Research, [grant number 16-32-00209].

**Figures (Color online)**

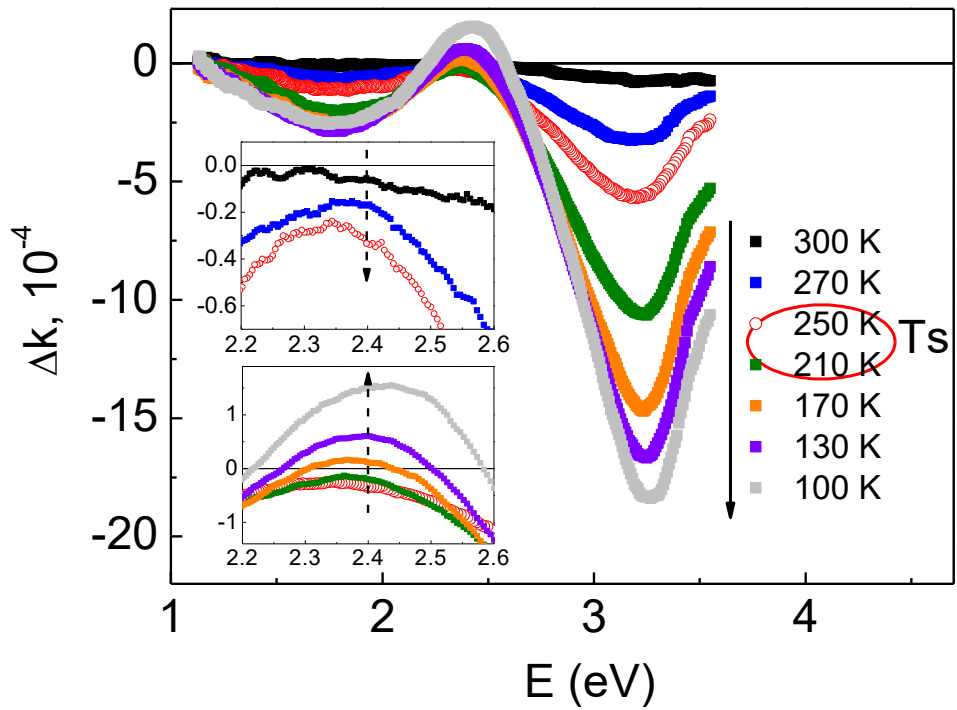

**Fig. 1.** MCD spectra at different temperatures for LSMO (x=0.3) film, d = 90 nm. Inserts: the scaled up regions between two main peaks in different temperature intervals: upper – $T > T_s$, lower – $T < T_s$. Magnetic field H = 3 kOe at oblique incidence of the light beam on the film plane ($12^0$).

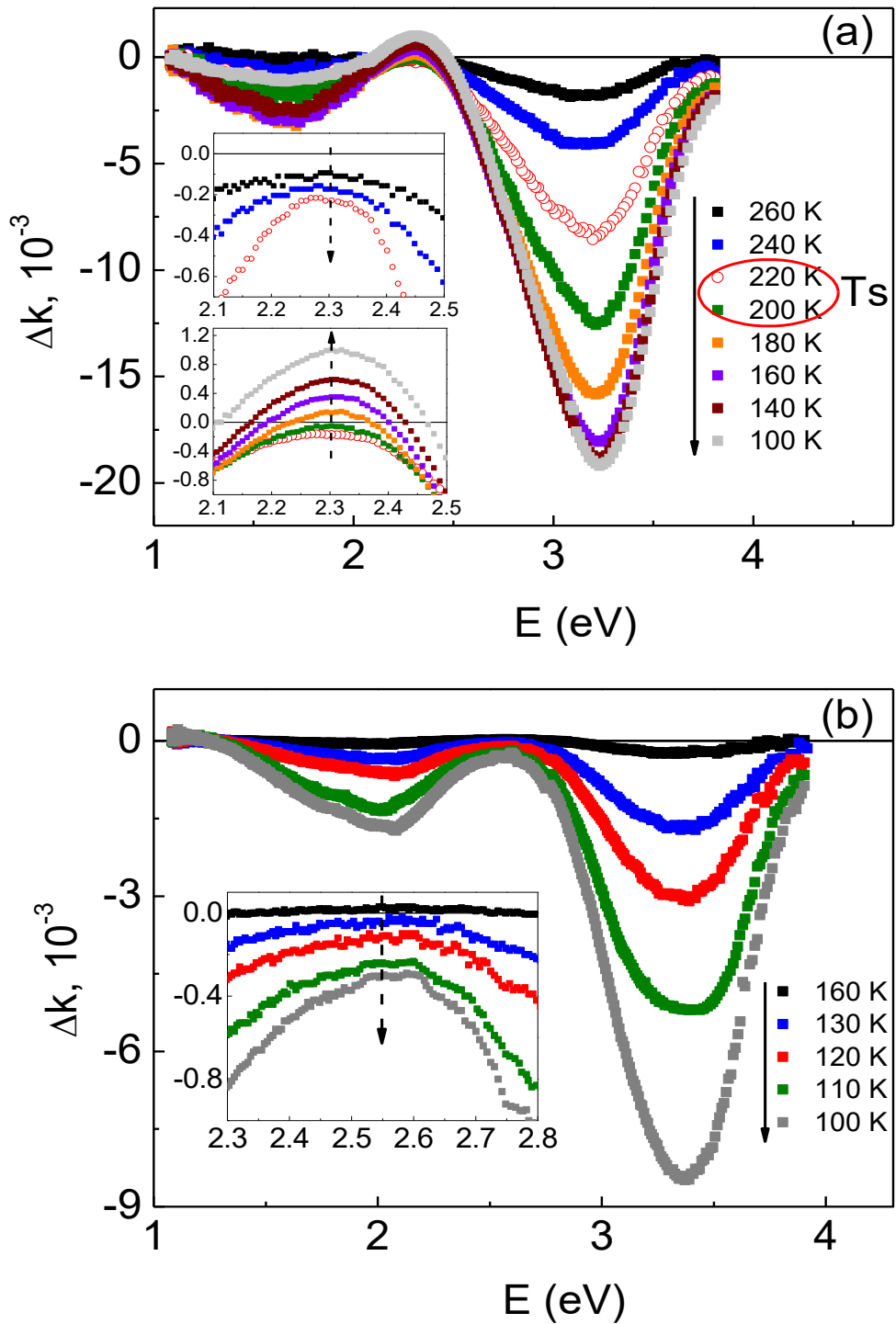

**Fig. 2.** MCD spectra at different temperatures: (a) PSMO (x = 0.4), d = 130 nm; (b) PSMO films (x = 0.2), d = 150 nm. Magnetic field H = 3.5 kOe applied normally to the film plane. Inserts: the scaled up regions between two main peaks at different temperatures.

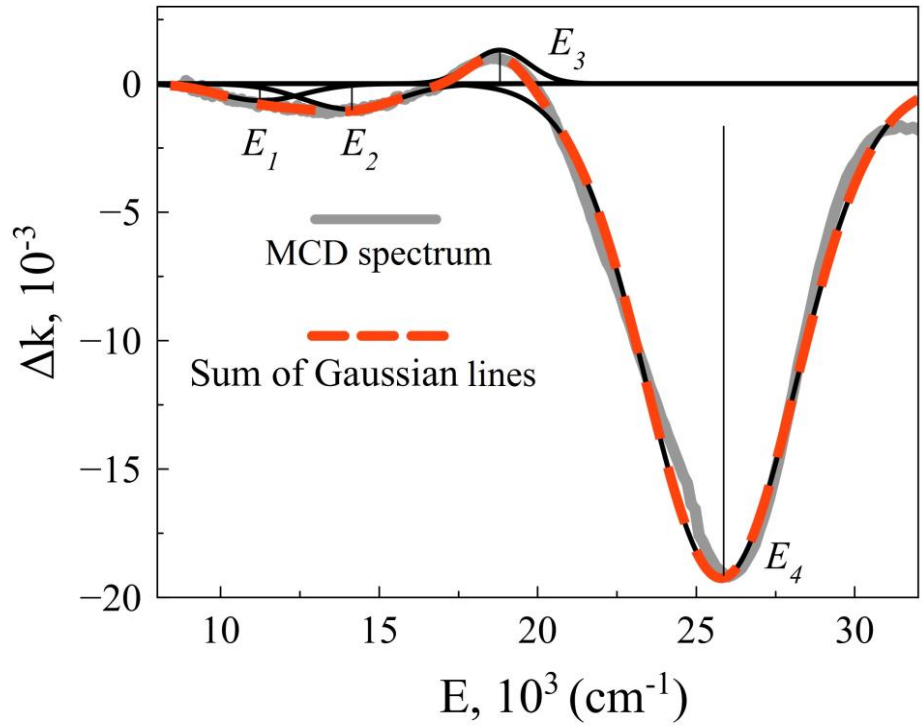

**Fig. 3.** Example of the MCD spectrum decomposition into Gaussian-shaped lines for the PSMO (x=0.4) sample at T = 100 K.

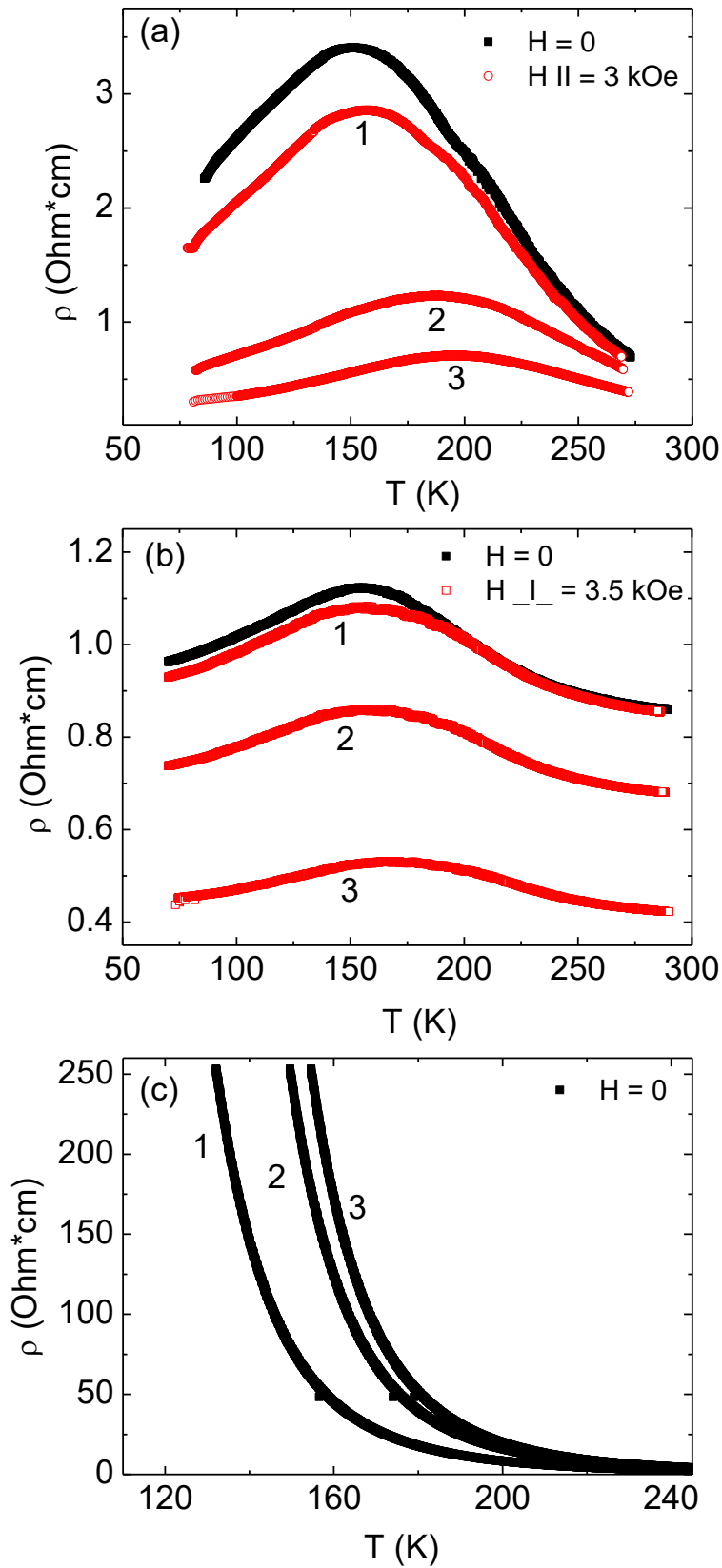

**Fig. 4.** Temperature dependences of resistivity of the manganite films without magnetic field (H = 0) and in magnetic field applied along ($H_{\parallel}$) and normally ($H_{\perp}$) to the sample plane: (a) LSMO (x=0.3), d = 20, 50, 90 nm (curves 1-3, respectively); (b) PSMO (x=0.4), d = 50, 80, 130 nm (curves 1-3, respectively); (c) PSMO (x=0.2), d = 50, 100, 150 nm (curves 1-3, respectively).

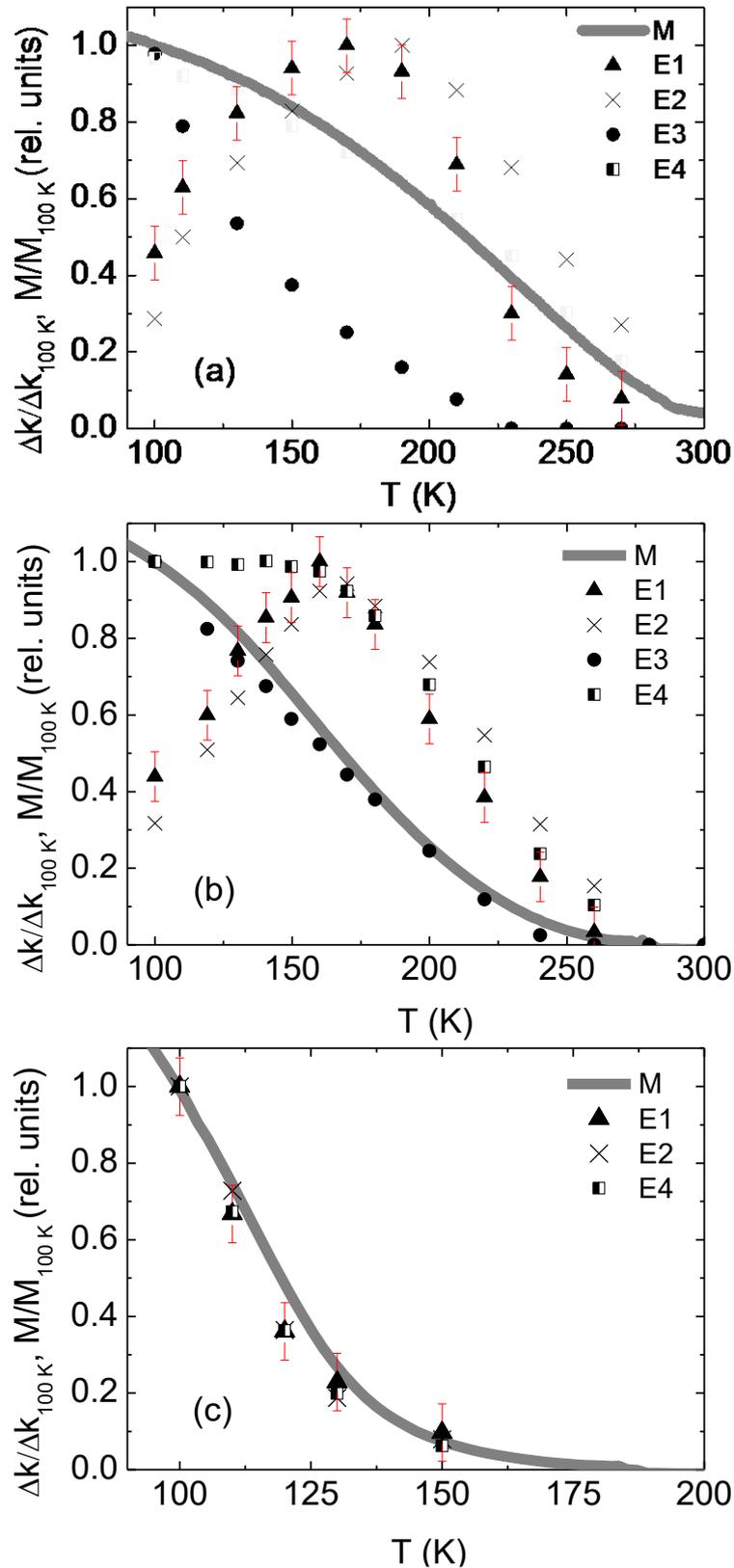

**Fig. 5.** Temperature dependences of the MCD bands intensity in comparison with the temperature behavior of the films magnetization (solid curves): (a) LSMO (x=0.3), d = 90 nm, H = 3 kOe; (b) PSMO (x=0.4), d = 130 nm, H = 3.5 kOe; (c) PSMO (x=0.2), d = 150 nm, H = 3.5 kOe. The temperature dependences of the samples magnetization were obtained at H = 3 kOe applied along the film plane (in the case of LSMO) and at H = 3.5 kOe applied normally to the film plane (in the case of PSMO). The bars show the error in determining the intensity of the bands.